# Does the Crab Have a Shell?

D. A. Frail

National Radio Astronomy Observatory, P.O. Box 0, Socorro, NM 87801

N. E. Kassim

Remote Sensing Division, Naval Research Laboratory, Code 7215, Washington, DC 20375-5321

T. J. Cornwell and W. M. Goss

National Radio Astronomy Observatory, P.O. Box 0, Socorro, NM 87801

## ABSTRACT

We present deep images of a region around the Crab nebula made with the VLA, utilizing new imaging and deconvolution algorithms in a search for a faint radio shell. The existence of a high-velocity, hydrogen-rich envelope has been predicted to account for the low total mass and kinetic energy of the observed nebula. No radio emission was detected from an extended source outside the Crab nebula. Our limits on surface brightness are sufficiently low to rule out the existence of a shell around the Crab whose brightness is at least two orders of magnitude below SN 1006, the faintest historical shell-type supernova remnant. We consider models for the progenitor star and the pre-supernova environment and conclude that if a fast, outer shock exists then it has a sharply reduced efficiency at accelerating relativistic particles from the kinetic energy of the blast wave. We also looked for a steepening of the spectral index along the boundary of the Crab nebula itself, the signature of an outer shock. However, contrary to previous claims, no such steepening was found. The absence of any evidence at radio wavelengths that either the Crab nebula or a hypothetical shell is interacting with the ambient medium leads to an interpretation that the supernova of 1054 AD was a peculiar low energy event.

*Subject headings:* supernova remnants — acceleration of particles — techniques: image processing

## 1. Introduction

The supernova (SN) of 1054 AD has long been regarded as a Type II event. Elemental abundances in the filaments constitute some of the strongest evidence in support for an SNII origin for the Crab (Henry & MacAlpine 1982, Davidson 1987), constraining the progenitor mass to lie in the range of 8-13 M⊙ (Nomoto 1985). However, the difficulty posed by this scenario is that the total *observable* mass (pulsar + filaments) can account for only 2.4-3.4 M⊙ (MacAlpine & Uomoto 1991). This low mass, combined with the low rms velocity of the filaments of only 1400 km s$^{-1}$ (Chevalier 1977) yields a kinetic energy for the SN1054 explosion of $2 \times 10^{49}$(M/M⊙) ergs – only a few percent of the canonical value of $10^{51}$ ergs for a Type II SN. This low value is particularly striking since it is an upper limit for the blast energy. The Crab's filaments are known to have been accelerated since 1054 AD (Trimble 1968, Wyckoff & Murray 1977) and it is plausible that the Crab pulsar is responsible for all or most of the total observed kinetic energy (Chevalier 1977).

Chevalier (1977, 1985) has suggested that this missing mass and energy may have been carried away in a high velocity, hydrogen-rich envelope from the outer layers of the progenitor star. Such shells have frequently been seen for historical and extragalactic SNIIs (Reynolds 1988, Bartel et al. 1994). However, despite repeated attempts at X-ray (Mauche & Gorenstein 1989, Predehl & Schmidtt 1995), optical (Murdin & Clark 1981, Murdin 1994) and radio wavelengths (Wilson & Weiler 1982, Velusamy 1984, Trushkin 1986, Velusamy & Roshi 1991) the shell around the Crab has eluded detection.

Whether the Crab has a shell or not at radio wavelengths has yet to be convincingly established. Particle acceleration occurring in the shocked transition zone between a shell and the ambient medium is expected to give rise to synchrotron radio emission. Existing limits are not good enough even to have detected a radio shell like the one seen for SN 1006, the faintest historical shell-type radio remnant. We present here high dynamic range images of the region around the Crab nebula, using the latest imaging and deconvolution algorithms, in a search for a faint radio shell.

## 2. Observations

The observations were made at the relatively low frequency of 333 MHz to reduce the brightness contrast between the flat-spectrum pulsar-powered nebula and the putative steep-spectrum shell. Two independent datasets were acquired. The first was made up of observations taken in the right hand of circular polarization in 6 sessions from 91 November to 93 March while the second has both hands of polarization taken in 94 September and 94 December. All data were acquired with the VLA in its B- and C-array configurations, giving a synthesized beam of 17″, a field-of-view (at half power) of ∼2° (156′), and a sensitivity to extended structure below ∼30′ in diameter. All observations used spectral line mode to enable excision of narrow-band interference (Cornwell, Uson & Haddad 1992).

High dynamic range imaging of the region around the Crab nebula is needed in order to search for a faint shell. In addition to careful editing of the data and self-calibration to correct for antenna-based gain errors (Perley 1989), there are other effects that must be compensated for before such an image can be made. Briggs (1995) has shown that in some circumstances the usual radio deconvolution algorithms (CLEAN and MEM: Cornwell & Braun 1989) can introduce errors that can equal or exceed the residual calibration errors in the data. Applying the normal deconvolution algorithms to the Crab nebula results in the introduction of non-physical structure beyond its boundaries. Self-calibration of the data can be improved by employing a non-negative least squares matrix inversion algorithm (NNLS). Instead of the usual iterative deconvolution process in the image plane, NNLS models the data via a direct algebraic deconvolution. Simulations show that NNLS is superior to CLEAN and MEM for sharp-edged sources like the Crab nebula (Briggs 1995).

NNLS produces a high-quality model of the Crab nebula necessary for calibration and sidelobe removal, but to achieve a noise limited image over the *entire* primary beam it is necessary to employ a 3D imaging algorithm. Except for snapshot observations, the VLA is a noncoplanar array. Therefore the usual 2D Fourier transform of the measured visibility data, which assumes a point spread function that does not vary across the field of view, introduces phase errors which increase with distance from the phase center. At low frequencies where the field of view is large these errors can become severe (Cornwell & Perley 1992). To overcome this and achieve noise limited images, first the NNLS model of the Crab nebula is subtracted from the original data, and then the sky is broken up into a series of overlapping regions (or facets), with each CLEAN'ed separately and combined in the final image. Both the algebraic deconvolution and the 3D imaging algorithm were carried out in the SDE package at NRAO, while all other processing used the AIPS package.

## 3. Results

A subsection of the wide-field image of the region around the Crab nebula is shown in Fig. 1 (Plate 1.). The angular extent and overall morphology is the same as that in previous radio images (e.g. Bietenholz & Kronberg 1990). The Crab's "jet" is clearly detected and unlike previous images (Velusamy 1984) it can be traced over the full extent of the [OIII] image by Fesen & Gull (1986). Outside the Crab nebula the rms fluctuations for the first dataset with a $20''$ synthesized beam are 1 mJy beam$^{-1}$, about 1.3 times the expected thermal noise. As the peak brightness in the nebula is approximately 10 Jy beam$^{-1}$ the dynamic range of this image is 10,000:1. The second dataset was contaminated by low level interference. The resulting rms noise is comparable to the first dataset but it is 3 times the expected thermal noise (in the absence of interference).

We find no evidence of extended emission outside the Crab nebula throughout the $2°$ field of view. For comparison, an unimpeded shock traveling at 10,000 km s$^{-1}$ for the lifetime of the Crab (940 yrs) would reach a diameter of $32'$. To further increase our sensitivity to extended emission the image was convolved with a circular Gaussian beam ranging from $0.5'$ to $2'$. Our surface brightness sensitivity $\Sigma$ is $7.5\times10^{-22}(1/\theta)$ W m$^{-2}$ Hz$^{-1}$ sr$^{-1}$, where $\theta$ is the size of the convolving beam in arcminutes. Previous searches for a radio shell around the Crab at 610 MHz and 1.4 GHz resulted in surface brightness limits of $2\times10^{-20}$ W m$^{-2}$ Hz$^{-1}$ sr$^{-1}$ (Wilson & Weiler 1982) and $6\times10^{-21}$ W m$^{-2}$ Hz$^{-1}$ sr$^{-1}$ (Velusamy 1984), respectively. Our present result improves on these limits by almost an order of magnitude.

Useful comparisons can also be made with known supernova remnant shells. In table 1 we list the properties of several shell remnants whose age can be determined from historical records or by a pulsar's spindown rate. For convenience all measured values of $\Sigma$ are scaled to 1 GHz, using a published spectral index or assuming a typical value of $\alpha = -0.5$ (where $\Sigma_\nu \propto \nu^\alpha$). In addition to the mean $\Sigma$ averaged over the *entire* diameter of the remnant we include estimates of $\Sigma_{\text{shell}}$, the mean surface brightness along the edge-brightened shell. While the former is often quoted for $\Sigma$, the latter is more relevant, resulting in a more realistic estimate of $\Sigma$ for comparison with our upper limits. If R is the radius of the shell and $\Delta$R is its width then $\Sigma = (2f - f^2)\Sigma_{\text{shell}}$, where $f=\Delta R/R$. For most supernova shells in table 1 $\Sigma_{\text{shell}}$ is 3-5 times larger than $\Sigma$, while for SN 1006 it is notably 14 times larger.

Like previous authors we can easily rule out a shell-type remnant around the Crab nebula whose $\Sigma$ is comparable to that of Cas A, Kepler or Tycho. For the first time this deep image enables a search to be made for a radio shell as faint as that in SN 1006. SN 1006 is a useful benchmark since, although it is thought to have been a SNI, it is similar in age to the Crab and it too was born at a high galactic z-height ($z_{crab} \sim 200$ pc versus $z_{SN1006} \sim 600$ pc). Similarly, SNR 0540−693 has been called the Crab's "twin" for both the pulsar and the remnant of each share many common physical properties (Reynolds 1985). A shell around the Crab with $\Sigma_{\text{shell}}$ equal to that of either SN 1006 or SNR 0540−693 would been 2-3 orders of magnitude brighter than our detection limits. There are old, large-diameter shell-type supernova remnants with low $\Sigma$ (e.g. G 156.2+5.7, G 65.2+5.7, OA 184 and S 147), but again, none of these has a $\Sigma_{\text{shell}}$ below our detection limit.

In summary, a deep, wide-field image around the Crab nebula has failed to detect a faint radio shell. The limits on the shell brightness are below that of all known historical supernova remnants. For the range of angular structures that the interferometer is sensitive to ($17''$-$30'$) not even the faintest cataloged supernova remnants in the Galaxy would have escaped detection.

## 4. Discussion

If there is a fast shell which carries away the bulk of the SN energy, then non-thermal radio emission may be expected to arise as it sweeps up the surrounding circumstellar or interstellar medium. Particle acceleration will occur either at the turbulent interface between the ejecta and the shocked ambient medium or at the outer shock itself (Ellison et al. 1994). Thus the absence

– 5 –of a radio shell around the Crab nebula, despite a deep search, has implications for the progenitor star and the pre-SN environment.

From standard synchrotron theory we can use our limit on $\Sigma_{shell}$ to estimate upper limits on the radio luminosity $L_R$ and minimum energy $E_{min}$ of the putative shell (Pacholczyk 1970). Taking the energy densities in relativistic particles and fields to be in equipartition and assuming reasonable shell parameters with $\Delta R/R=0.15$, $\alpha = -0.5$ (from $10^7$-$10^{11}$ Hz) and an energy ratio of protons to electrons accelerated in the shock (the k-factor) of 40, $L_R \leq 2.2 \times 10^{31} (R_s/10 \text{ pc})^2$ erg s$^{-1}$ and $E_{min} \leq 6.2 \times 10^{47} (R_s/10 \text{ pc})^{17/7}$, where $R_s$ is the shell radius.

The radio surface brightness of the putative shell and its luminosity depends in part on the ambient density into which the SN shock expands. Chevalier (1985) has examined the evolution of massive stars with an emphasis on how they modify their pre-SN environment. In the blue giant phase a hot, shocked wind creates a low density cavity out to a radius of several tens of pc. Romani et al. (1990) and Wallace, Landecker & Taylor (1994) present evidence that SN 1054 exploded into a low density region. Using low resolution observations of the 21-cm line of atomic hydrogen (HI) they find that the Crab nebula sits near a local HI minimum, around which there is a partial shell ($R \simeq 70$ pc) of swept-up gas ($M_{tot} \simeq 2 \times 10^4$ M$\odot$). They argue that this is not a blue giant bubble but a pre-existing cavity formed by the action of multiple supernova remnants or by stellar winds from a cluster of massive stars. For the age of the Crab a blast wave expanding in this low density $n_o$ medium will have a radius given by $R_s=7.8 (E_o/10^{51} \text{ ergs})^{2/7} (M_{ej}/5 \text{ M}\odot)^{-1/7} (n_o/0.01 \text{ cm}^{-3})^{-1/7}$ pc, where $E_o$ is the total kinetic energy, and $M_{ej}$ is the mass of the ejecta (Chevalier 1985). The outer shock expands at a mean velocity of 8100 km s$^{-1}$, sweeping up $0.7(E_o/10^{51} \text{ ergs})^{6/7}(M_{ej}/5 \text{ M}\odot)^{-3/7}(n_o/0.01 \text{ cm}^{-3})^{4/7}$ M$\odot$ of the ambient gas.

At a radius $R_s$ the efficiency $\epsilon$ of the shock in converting its kinetic energy into synchrotron particles and magnetic field is given by $E_{min}/E_o$. For the parameters assumed above (i.e. $E_o=10^{51}$ ergs, $M_{ej}=5$ M$\odot$, $n_o/0.01$ cm$^{-3}$) $\epsilon=3\times 10^{-4}$. This is sharply lower than typical values. The identical analysis applied to the remnants in table 1 yields values of $\epsilon$ of 0.003 to 0.05. In a large sample of supernova remnants in M33 Duric et al. (1995) derive $\epsilon=0.01$–0.1. Our limit on $\epsilon$ is more severe for larger $n_o$ (and the swept-up mass is larger). For reference, X-ray observations of SN 1006 and G 156.2+5.7 (the faintest known radio supernova remnant) give $n_o$ of 0.1 cm$^{-3}$ and 0.01 cm$^{-3}$, respectively (Jones & Pye 1989, Pfeffermann et al. 1991). We conclude that if the Crab's outer shock exists it has a much reduced efficiency for accelerating relativistic particles.

On the other hand there is evidence that the Crab's progenitor star may have had substantial mass loss prior to the SN event. If the massive progenitor star lost all or most of the hydrogen envelope during its red giant phase it removes the necessity for a fast outer envelope. The "bays" or indentations in the optical synchrotron nebula have long been known to exist. These have been interpreted as part of a torus, resulting either from mass loss in a thin disk (Fesen, Martin & Shull 1992), or an equatorially-concentrated wind (Li & Begelman 1992) of a massive star in the red giant phase. Murdin (1994) has put forth a similar explanation for the origin of the faint H$\alpha$ halo around the Crab, first detected by Murdin & Clark (1981). However, the existence of this H$\alpha$ halo remains controversial since it has yet to be confirmed by other investigators (e.g. Gull & Fesen 1982) and its inferred density is above the upper limits deduced by the absence of deceleration for the synchrotron nebula (Bietenholz et al. 1991). Although we recognize that the evidence for a dense outer envelope is far from compelling we now consider the effects of such a medium on the radio properties of the nebula.

A shock will form in the dense, slow moving wind due to the interaction of either the hypothesized fast outer shell or the Crab nebula itself (Chevalier 1984). Distinguishing this feature from the body of the nebula at radio wavelengths may be possible since its spectrum should be characteristic of shock acceleration ($\alpha \simeq -0.5$) rather than acceleration in the pulsar magnetosphere ($\alpha_{crab} \simeq -0.26$). Evidence for a steepening of $\alpha$ for the emission at the outer boundaries of the nebula has been given by Velusamy (1985) and Velusamy, Roshi & Venugopal (1992). With better quality data Bietenholz & Kronberg (1992) show that any systematic steepening in $\alpha$ towards the edge of the nebula is $< 0.01$ (see also Trushkin 1986). A comparison of our 333 MHz image with the 1.5 GHz image of Bietenholz & Kronberg (1992) confirms these findings. A high resolution spectral index study of the Crab and its jet will be reported in a later paper (Kassim, Bietenholz & Frail, in preparation). Similarly, there is no indication of a steep spectrum shell in the *integrated* spectrum of the Crab nebula. After subtracting the flat spectrum component of the Crab nebula from low frequency data from 10 MHz to 1 GHz compiled by Kovalenko, Pynzar' & Udal'tsov (1994) we find that the maximum residual flux that can be fit to an $\alpha = -0.5$ slope is 18 Jy at 100 MHz. We estimate that the upper limit on the radio luminosity of any shock accelerated emission is $L_R = 5.4 \times 10^{32}$ erg s$^{-1}$, only 0.3% of the total radio luminosity of the Crab nebula. Thus any shock at the edge of the Crab is energetically unimportant and remains hidden by the pulsar-powered synchrotron nebula.

In summary, we find no evidence for a fast outer synchrotron shock around the Crab nebula, nor does there appear to be any evidence of an interaction with the ambient medium in the immediate vicinity of the Crab nebula. The wide spread in the ratio of $L_R$ (or $\Sigma$) between the shell and pulsar-powered components of composite remnants (Helfand & Becker 1987) has been attributed to remnants in different stages of evolution, each with a range of initial conditions, including $n_o$, $E_o$, the shock velocity, initial pulsar period and magnetic field (Bhattacharya 1990). However, the Crab now joins other pulsar-powered nebulae, including 3C 58 (Reynolds & Aller 1985), for which deep imaging has failed to find evidence of a radio shell. The problem is more acute for G 74.9+1.2, another pulsar-powered nebula which is also at a large z-distance (z$\sim$ 250 pc) but which is at least three times as old as the Crab. Wallace et al. (1995) present compelling evidence that G 74.9+1.2 is adjacent to an HI shell and is interacting with its inner boundary. Like the Crab, the absence of an edge-brightened shell, despite clear signs that the remnant is interacting with the ambient medium, raises the possibility that no fast envelope was expelled at the time of the SN event. Perhaps the SN of 1054 AD was a peculiar low energy event ($E_o << 10^{51}$ ergs). Nomoto (1985, 1987) has been the leading advocate of this interpretation and has suggested several possible progenitor models. Pols (1994) has devised an interesting variant on these models in which the primary sheds its H-rich envelope onto the secondary of a wide binary system and the secondary explodes first. The primary, now a He star, is ejected from the system at high velocity. Such a model can nicely explain the high z of the Crab, the He-rich filaments, the absence of a H envelope and the lack of an obvious companion.

The Very Large Array (VLA) is a facility of the National Science Foundation operated under cooperative agreement by Associated Universities, Inc. DAF thanks J. Uson for pushing the development of P-band capabilities at the VLA, without which this experiment would not have been possible. Basic research in radio astronomy at the Naval Research Laboratory is supported by the Office of Naval Research. This research has made use of the Simbad database, operated at CDS, Strasbourg, France.

Table 1. Young Supernova Remnants.

| Name | Age (yrs) | Shell Diameter (pc) | Distance (kpc) | $\Sigma$ (W m$^{-2}$ Hz$^{-1}$ sr$^{-1}$) | $\Sigma_{\text{shell}}$ (W m$^{-2}$ Hz$^{-1}$ sr$^{-1}$) |
|---|---|---|---|---|---|
| Cas A (SN 1680) | 314 | 4.1 | 2.8 | $1.6 \times 10^{-17}$ | $6.5 \times 10^{-17}$ |
| Kepler (SN 1604) | 390 | 3.8 | 4.4 | $3.2 \times 10^{-19}$ | $9.0 \times 10^{-19}$ |
| Tycho (SN 1572) | 422 | 5.4 | 2.3 | $1.3 \times 10^{-19}$ | $2.8 \times 10^{-19}$ |
| SNR 0540−693 | 760 | 17 | 55 | $1.1 \times 10^{-19}$ | $5.4 \times 10^{-19}$ |
| SN 1006 | 988 | 21 | 2.4 | $3.2 \times 10^{-21}$ | $4.5 \times 10^{-20}$ |
| Crab (SN 1054) | 940 | | 2.0 | $< 4.3 \times 10^{-22}$ | $< 4.3 \times 10^{-22}$ |

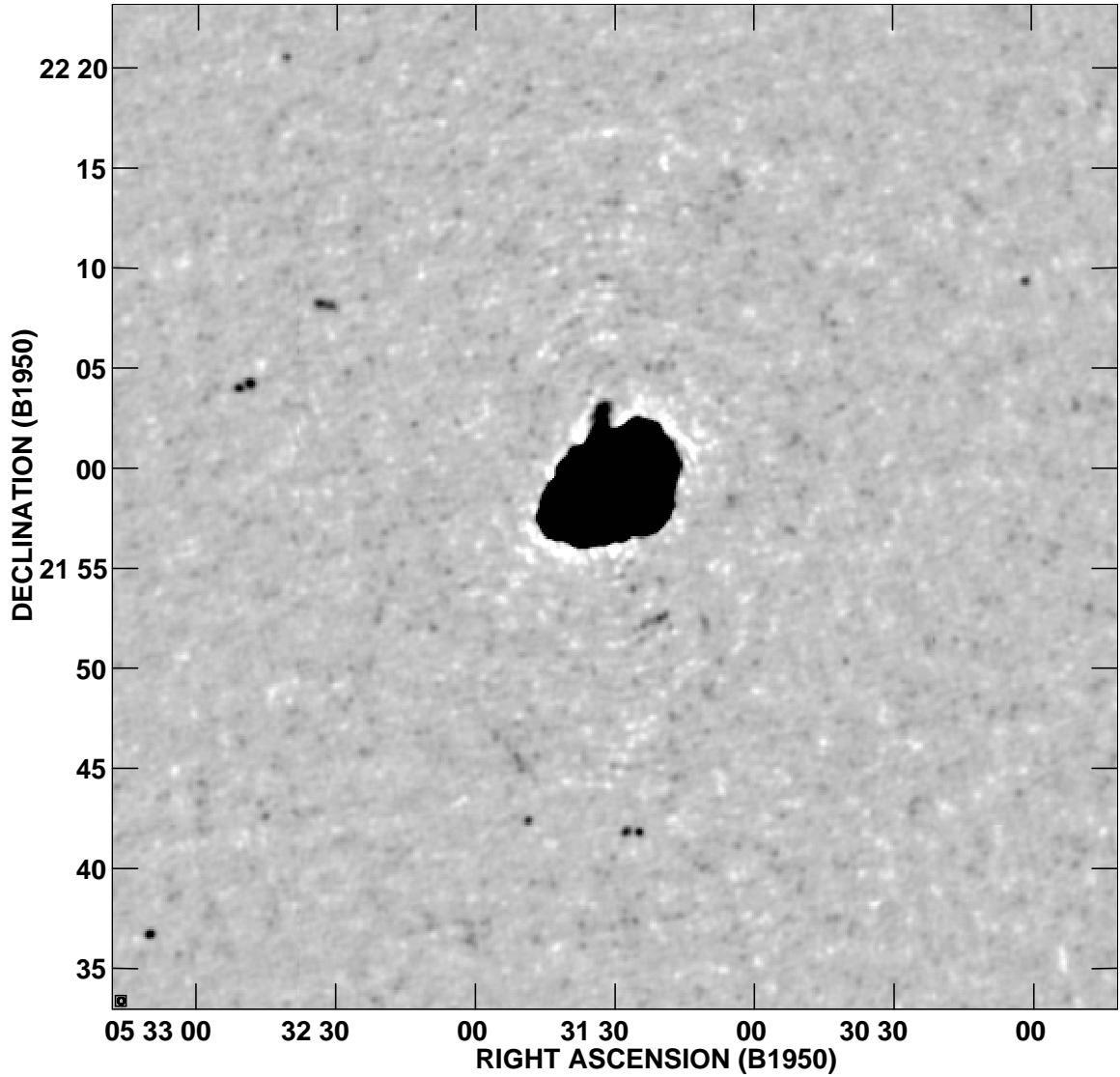

Fig. 1.— A radio continuum image of the field around the Crab nebula taken with the VLA at 333 MHz. The synthesized beam is $20''$ and the rms noise on the image is 1 mJy beam$^{-1}$. Greyscale levels are set to emphasize emission from $-10$ mJy beam$^{-1}$ to $+30$ mJy beam$^{-1}$. The peak flux density on the Crab nebula is 12 Jy beam$^{-1}$. The negative emission near the outside edge of the Crab nebula is due to residual calibration errors not fully removed by the deconvolution algorithms.